\documentclass[prd,showpacs,showkeys,superscriptaddress]{revtex4}
\usepackage{amsmath}
\usepackage[pdftex]{graphicx}
\usepackage[english]{babel}

\begin{document}

\title{Thermodynamical analysis on a braneworld scenario with curvature corrections}
 \author{Ninfa Radicella}
 \email{ninfa.radicella@uab.cat}
  \author{Diego Pav\'on}
 \email{diego.pavon@uab.es} 
\affiliation{Departamento de F\'isica, Universidad Aut\'onoma de
Barcelona, 08193 Bellaterra (Barcelona), Spain}

\begin{abstract}
We study the thermodynamics of some cosmological models based on modified gravity, a braneworld with induced gravity and curvature effect. Dark energy component seems necessary if the models are to approach thermal equilibrium in the long run.
\end{abstract}
\maketitle

\section{Introduction}
Supernoave type Ia data \cite{riess98} as well as other observational probes \cite{komatsu10} show that the Universe is undergoing an accelerated phase of expansion at present time, a feature that does not emerge from standard cold dark matter model \cite{peebles93}. Attempts to explain such unexpected behaviour go towards modifications of either the geometric part of the Einstein field equations, implying modified theories of gravity, or the matter sector, thus involving new and sometimes weird forms of energy \cite{reviews}. Such a dark sector seems unavoidable in order to fit present cosmological data.\\
On the other hand, from the theoretical side, the strong mathematical resemblance between the dynamics of spacetime horizons and thermodynamics is strongly attested \cite{jacobson95, padmanabhan05} so that gravitational fields equations can be given a physical interpretation which is thermodynamical in origin. In particular, the Friedmann equations follow from applying the first law to the apparent horizon of an isotropic and homogeneous universe, not only in Einstein gravity, but also in more general Lovelock gravity \cite{cai05}. Likewise, it seems that a gravitational theory built on the principle of equivalence must be thought of as a macroscopic limit of some underlying microscopic theory, the microscopic structure of spacetime manifesting itself only at Planck scale or near singularities. Also the horizons link some aspects of microscopic physics with the bulk dynamics \cite{padmanabhan98}. It is well known since long that one can define entropy and temperature for a spacetime horizon \cite{hawking75, unruh76, gibbons77}; in fact, many attempts have been done to better understand this link. An instructive example is the case of spherically symmetric horizons in four dimensions, for which Einstein's equations can be interpreted as a thermodynamic relation arising out of virtual displacement of the horizon \cite{padmanabhan02}. Moreover, the same interpretation holds for the case of the Lanczos-Lovelock gravitational theory in D dimensions \cite{paranjape06} and explicit demonstration has been given for Friedmann models \cite{cai07a}.

In the present paper, following this deep relation between thermodynamic and gravity, and in particular between entropy and horizons, we argue that some form of dark energy is demanded on thermodynamic grounds.\\
 In order  for an isolated system to evolve to thermodynamical equilibrium, the entropy function of the system must show two properties: it must never decrease, i.e. its first derivative with respect to  the relevant variable must be non-negative, and convex, i.e., its second derivative must be negative.\\ This constitutes  the hard core of the second law of thermodynamics and it is naturally realised in systems dominated by electromagnetic forces; however it might not be true when gravity plays a role. In fact, the entropy of the system must still increase but it may be grow unbounded: this occurs, in the Newtonian framework, for the Antonov's sphere, the final stage of $N$ gravitating point masses enclosed in a perfectly reflecting, rigid, sphere whose radius exceeds some critical value \cite{antonov62,lynden68}. Nevertheless, when we replace Newtonian gravity by general relativity, a black hole is expected to be formed at the center of the sphere that, though it tends to evaporate, it will likely arrive to an equilibrium state characterised by a state of maximum, finite, entropy. \\
In any case, in a Friedmann-Robertson-Walker (FRW) cosmology, the Universe seems to behave as an ordinary system whose entropy increases towards a maximum value. The latter follows from the observational data on the evolution of the Hubble factor of the FRW metric \cite{pavon11} and from the evolution of the entropy of the apparent horizon, that seems to be the appropriate thermodynamic boundary \cite{wang06}. 
 
The present paper is a second step of the analysis outlined above. Infact, in a previous paper \cite{radicella10} we showed that an Einstein Universe, as a thermodynamical system, cannot tend to equilibrium in the last stage of expansion unless it accelerates. We have found that this holds true for some modified models that are dynamically equivalent at the background level, nevertheless this does not mean that every accelerating universe is thermodynamically motivated \cite{radicella10}.\\ In this work we study the thermodynamical behaviour of a braneworld model with two correction terms: a four-dimensional curvature on the brane and a Gauss-Bonnet (GB) term in the bulk \cite{kofinas03}. The induced gravity (IG) correction arises because the localized matter fields on the brane, which couple to bulk gravitons, can generate via quantum loops a localized four-dimensional world-volume kinetic term for gravitons \cite{dvali01}. On the other hand, a Gauss-Bonnet term naturally appears in an effective action approach to string theory, corresponding to the leading order quantum corrections to gravity \cite{zwiebach85}. As a result, we have the most general action with second-order field equations in five dimensions \cite{lovelock71}. 
 
 Section \ref{GBIG} introduces the braneworld cosmology with induced gravity and curvature effects. Subsections \ref{hor} and \ref{mat} focus on the entropy of the horizon and matter components, respectively. The energy componentsof the Universe are assumed to enter the field equations in the form of perfect fluids, the standard equation of state being true for each of them: $p_i=w_i \rho_i$. Then, in section \ref{chaplygin}, the matter component is assumed as cold matter and a Chaplygin gas. The choice of a Chaplygin gas is based on the recent observational result that the equation of state parameter of dark energy can be less than $-1$ and even display a transient behaviour \cite{komatsu10}. This can be achieved either by means of phantom fields, that on the other hand suffer from instabilities \cite{cline04}, or by other approaches that mimic this phantom-like behaviour.  In the model under analysis, in which UV modifications are included by considering the stringy effect via the GB term in the bulk, and IR modifications are due to the IG effect, a Chaplygin gas fluid on the brane provides a smooth crossing of the cosmological constant line \cite{nozari11}. In fact, this component is characterised by a cross-over length scale below which the gas behaves as pressureless fluid and above which it mimics a cosmological constant.\\
 Our conclusions are given in section \ref{conclusions}: we find that even this modified theory of gravity needs a component with typical dark energy behaviour in order to satisfy the generalised second law (GSL) and approach thermodynamical equilibrium in the long run.


\section{Gauss-Bonnet and Induced Gravity corrections}\label{GBIG}
The total action of the braneworld model under consideration reads \cite{kofinas03}
\begin{equation}
I=\frac{1}{2 \kappa_5^2}\int d^5 x \ \sqrt{-^{(5)}g}\left(^{(5)} R-2\Lambda_5+\alpha \mathcal{L}_{GB}\right)+\frac{1}{2 \kappa_4^2}\int d^4 x\ \sqrt{^{(4)}g} \left(R - 2\Lambda_4\right)+\int d^4 x\  \mathcal{L}_m,
\end{equation}
where $\Lambda_5<0$ is the cosmological constant on the bulk and 
\begin{equation}
\mathcal{L}_{GB}=\ ^{(5)}R^2-4\ ^{(5)}R^{AB}\ ^{(5)}R_{AB}+ ^{(5)}R^{ABCD} \ ^{(5)}R_{ABCD}
\end{equation}
is the GB correction term, whose coupling constant $\alpha=1/8 g_s^2$ is related to the string energy scale, $g_s$ \cite{kofinas03}.  The gravitational coupling constants $\kappa_4^2=8\pi G_4$ and  $\kappa_5^2=8\pi G_5$ on the brane and in the bulk, respectively introduce a length scale, the induced gravity cross-over scale, $r=\kappa_5^2/2\kappa_4^2$ and help defining the brane tension, $\lambda=\Lambda_4/\kappa_4^2$. Last term represents matter action.

\subsection{Cosmological equations}
The metric of homogeneous and isotropic FRW Universe on the brane, with spatial curvature index $k$, is
\begin{equation}
ds^2=h_{\mu\nu}dx^\mu dx^\nu +\tilde{r}^2\left[d\theta^2+\sin{\theta}^2 d\phi^2\right],
\end{equation}
where $\tilde{r}=a(t)r$, the two-dimensional metric is $h_{\mu\nu}=\text{diag}\left(-1,a^2/(1-kr^2)\right)$ with $x^0=t$ and $x^1=r$. This allows the explicit evaluation of the radius of the apparent horizon  (a marginally trapped surface with vanishing expansion) determined by the relation $h^{\mu\nu}\partial_{\mu}\tilde{r}\partial_{\nu}\tilde{r}=0$ \cite{bak00} that gives 
\begin{equation}
\tilde{r}_A=\frac{1}{\sqrt{H^2+\frac{k}{a^2}}}.
\end{equation}
Friedmann's equation on the brane is 
\begin{equation}\label{friedmann}
-\frac{1}{r}\left[1+\frac{8}{3}\alpha \left(H^2+\frac{k}{a^2}+\frac{\Phi_0}{2}\right)\right]\left(H^2+\frac{k}{a^2}-\frac{\Phi_0}{2}\right)^{1/2}=-\frac{k_4^2}{3}(\sum_i\rho_i+\lambda)+H^2+\frac{k}{a^2},
\end{equation}
in which, assuming there is no black hole in the bulk, $\Phi_0=\frac{1}{4\alpha}\left(-1+\sqrt{1+\frac{4}{3}\alpha\Lambda_5}\right)$ and matter field are supposed to be perfect fluids with energy density $\rho_i$ so that in order to describe completely the cosmological dynamics on the brane we can use the energy conservation law
\begin{equation}\label{cons}
\dot{\rho_i}+3H\rho_i(1+w_i)=0,
\end{equation}
where $w_i=p_i/\rho_i=\text{const}$.
\subsection{Entropy of the apparent horizon}\label{hor}
The entropy on the apparent horizon is given by \cite{cai07b, sheykhi07, sheykhi09}:
\begin{equation}
S_A=4\pi\left[\frac{1}{2 G_4}\int_0^{\tilde{r}_A}\tilde{r}_A d\tilde{r}_A+\frac{1}{2 G_5}\int_0^{\tilde{r}_A}\frac{\tilde{r}^2_A d\tilde{r}_A}{\sqrt{1-\Phi_0\tilde{r}_A^2}}+\frac{2 \alpha}{G_5}\int_0^{\tilde{r}_A}\frac{2-\Phi_0 \tilde{r}^2_A }{\sqrt{1-\Phi_0\tilde{r}_A^2}} d\tilde{r}_A\right].
\end{equation}
Its derivatives with respect to the scale factor $a$, that will be denoted by a prime, can be calculated and simplified by using eqs.(\ref{friedmann})-(\ref{cons})
\begin{eqnarray}
S'_A&=&8\pi^2 \tilde{r}^4_A\frac{\sum_i(\rho_i+p_i)}{a}=8\pi^2 \tilde{r}^4_A\frac{\rho_{T}(1+w_{T})}{a},\label{Sdothorizon}\\
S''_A&=&\frac{S'}{a}\left[4\frac{\tilde{r}'_A a}{r_A}-(3 w_{T}+4)\right],\label{Sddothorizon}
\end{eqnarray}
where $w_{T}=p_{T}/\rho_{T}=\sum_i p_i/\sum_i \rho_i$.

It can be easily checked that the entropy grows provided the fluids component of the Universe satisfy $w_T>-1$. In order to evaluate the second derivative we make use of the late time behaviour of this brane cosmology with curvature correction that, as found in \cite{kofinas03}, reduces to conventional cosmology with positive effective gravitational and cosmological constants. For $a\rightarrow\infty$ eq.(\ref{friedmann}) reduces to
\begin{equation}
H^2+\frac{k}{a^2}\approx G_{eff}\sum_i\rho_i+\Lambda_{eff},
\end{equation}
with
\begin{equation}
G_{eff}=3\nu G_4\quad\quad\text{and}\quad \quad \Lambda_{eff}=\frac{4-3\beta-\gamma}{4\beta\alpha},
\end{equation}
$\beta$, $\gamma$ and $\nu$ being functions of the parameters of the model, namely $\alpha$, $\Lambda_4$, $\Lambda_5$ and $r$, - see \cite{kofinas03}.
By using this solution the second derivative of the horizon entropy w.r.t. the scale factor, it gives
\begin{equation}
S''_A\approx-(4+3w_T) a^{-(5+3w_l)},
\end{equation}
where $w_l$ is the equation of state parameter of the fluid that redshifts more slowly. Thus, entropy tends to equilibrium  if $w_T>-1$, which is required by the GSL.
\subsection{Entropy of matter fields}\label{mat}
We first consider a perfect fluid with $w_i\neq0$. In this case, using Gibbs' law \cite{callen85}
\begin{equation}
T_f dS_f=d(\rho_i V)+p_i dV,
\end{equation}
with $V=4\pi \tilde{r}_A^3/3$. The evolution of the fluid temperature is governed by $d \ln{T_f} /d\ln{a} =- 3w_f $, see e.g. \cite{calvao92},
whence $T_f = T_{f0} a^{-3w_i}$. This means that fluids with $w_f>0$ dilute as the universe expands and the temperature decreases while dark energy fields behave the opposite and the temperature will grow during the expansion; for details see \cite{pavon09}. We then compute the two first derivatives of the entropy of the fluid enclosed by the apparent horizon:
\begin{eqnarray}
S'_{fi}&=&4\pi\frac{ \tilde{r}_A^3\rho_i(1+w_i)}{T_{fi}a}\left[a\frac{\tilde{r}'_A}{\tilde{r}_A}-1\right]\label{Sdotfluid}\\
S''_{fi}&=&4\pi\frac{ \tilde{r}_A^2\rho_i(1+w_i)}{T_{fi} a}\left[a \tilde{r}''_A+2a\frac{\tilde{r}'^2_A}{\tilde{r}_A}-3\tilde{r}'_A(2+w_i)+\frac{\tilde{r}_A}{a}(4+3w_i)+\frac{T'_{fi}}{T_{fi}}\left(\tilde{r}_A-a\tilde{r}'_A\right)\right]\label{Sddotfluid}.
\end{eqnarray}
A simple evaluation of the first derivative  for $a\rightarrow\infty$ shows that $S'_f\approx -a^{-4}$ when $1+w_i>0$: it tends to zero from below. To discern which component prevails in the long run we evaluate the ratio between the derivatives:
\begin{equation}
\frac{S'_f}{S'_A}=\frac{(1+w_i)}{2\pi(1+w_T) \tilde{r}_A}\ \frac{\rho_i}{\rho_TT_f}\left[a\frac{\tilde{r}'_A}{\tilde{r}_A}-1\right]\sim a^{3 w_l}
\end{equation}
provided $-1<w_l<0$. The same holds true for the ratio $S''_f/S''_A$.

Last section is devoted to dust fluid, for which we cannot use the evolution of temperature in terms of the scale factor as above because it gives the unphysical result of a vanishing constant temperature. Then, we proceed as in \cite{radicella10} in which every dust particle contributes a given bit to the fluid entropy so that $S_m=k_B$ N with $N=4\pi \tilde{r}^3_A n/3$ being the number of particles within the apparent horizon and $n=n_0 a^{-3}$ the number density of dust particles. \\
By computing first and second derivatives of dust entropy inside the apparent horizon we find
\begin{equation}\label{Sdotdust}
S'_m=\frac{4\pi k_B n_0 r^3_A}{a^4}\left(a\frac{\tilde{r}'_A}{\tilde{r}_A}-1\right),
\end{equation}
and
\begin{equation}\label{Sddotdust}
S''_m=\frac{4\pi k_B n_0 r^2_A}{a^4}\left(a \tilde{r}''_A -\frac{a \tilde{r}'^2_A}{\tilde{r}_A}+4\frac{\tilde{r}_A}{a}\right).
\end{equation}
Now, by computing the ratio between the components as before, we obtain
$$
\frac{S'_m}{S'_A}\sim\frac{S''_m}{S''_A}\sim a^{3 w_l}.
$$
It means that if the only fluid component is dust, we need to compute explicitly the sum of the horizon and matter contribution to the entropy derivatives. By restoring proper dimensions  we get
\begin{equation}\label{GSL}
S'_m+S'_A=4\pi k_B\frac{\tilde{r}^3_A}{a^4}\left[-n_0+\frac{\tilde{r}_A\rho_{m0}}{\hbar c}\right]
\end{equation}
and
\begin{equation}\label{equilibrium}
S''_m+S''_A=16\pi k_B\frac{\tilde{r}^3_A}{a^5}\left[n_0-\frac{ \tilde{r}_A\rho_{m0}}{\hbar c}\right].
\end{equation}
In order to fulfill the GSL the right hand side of eq.(\ref{GSL}) must be always positive and to approach the equilibrium eq.(\ref{equilibrium}) must be negative in the last stage of expansion. These two requirements translate in the following constraint 
\begin{equation}
\tilde{r}_A>\frac{\hbar c n_0}{\rho_{m0}}.
\end{equation}
Now, since $\rho_{m0}=n_0 m c^2$, where $m$ is the mass of the particles of the cosmological fluid, the apparent horizon radius must be larger than  the associated reduced Compton wavelength of the particles, $\lambda_C=\hbar /m c$.

If, on the contrary, we had more components, $w_l>0$ would imply that the dust term would dominate in the long run and its derivatives fulfill neither the GSL nor the equilibrium prescription. By requiring $-1<w_l<0$ the horizon entropy will prevail in the long run and the Universe tend to equilibrium, as the analysis of the previous section shows.

To sum up, a dark energy component (i.e.,  a barotropic fluid with $-1<w<0$) in the late time evolution is still needed, in such a modified theory of gravity, if the Universe is to tend to equilibrium eventually.

\section{A dark energy component: Chaplygin gas}\label{chaplygin}
As concluded is the previous section, it seems that from a thermodynamical point of view a dark energy component is needed. 
In what follows we restrict ourselves to a Minkowski bulk and a spatially flat FRW brane so that the radius of the apparent horizon now reduces to $\tilde{r}_A=1/H$. When $\Lambda_5=0$, $\Phi_0$ could be zero or $\Phi_0=-1/2\alpha$; we use the first choice since the latter case has no IG limit: it corresponds to an AdS bulk even though $\Lambda_5=0$ \cite{brown05}.
Cosmological evolution still follows eqs. (\ref{friedmann})-(\ref{cons}) where the total energy density is now
\begin{equation}
\rho=\rho_m+\rho_{ch}
\end{equation}
and the CDM and Chaplygin gas components obey:
\begin{eqnarray}\label{conschap}
\dot{\rho}_m&+&3H\rho_m=0,\\
\dot{\rho}_{ch}&+&3H(\rho_{ch}+p_{ch})=0,
\end{eqnarray}
respectively.
Then CDM redshifts as $\rho_m=\rho_{m0}a^{-3}$ while, being the Chaplygin pressure $p_{ch}=w_{ch}\rho_{ch}=-A\rho_{ch}^{-\tilde{\alpha}}$, 
\begin{equation}\label{chapevol}
\rho_{ch}=\rho_{ch0}\left[A_s+\frac{1-A_s}{a^{3(1+\tilde{\alpha})}}\right]^{\frac{1}{1+\tilde{\alpha}}},
\end{equation}
where $A_s=A \rho_{ch0}^{-(1+\tilde{\alpha})}$, and $\tilde{alpha}$ a constant.
\subsection{Entropy of the apparent horizon}
The first and second derivatives of the apparent horizon entropy $S_A$ for such a model read
\begin{eqnarray}
S'_A&=&8\pi^2 \tilde{r}^4_A\frac{\sum_i(\rho_i+p_i)}{a},\label{Sdothorchap}\\
S''_A&=&\frac{S'_{A}}{a}\left[4\frac{\tilde{r}'_A a}{r_A}-4-3 A \tilde{\alpha} \frac{(\rho_{ch}+p_{ch})\rho_{ch}^{-(\tilde{\alpha}+1)}}{\rho_m+\rho_{ch}+p_{ch}}\right].\label{Sddothorchap}
\end{eqnarray}

\begin{figure*}[htb]
   \hspace{-3.5 cm}
\begin{minipage}{0.3\textwidth}
\centering
 \includegraphics[width=8cm]{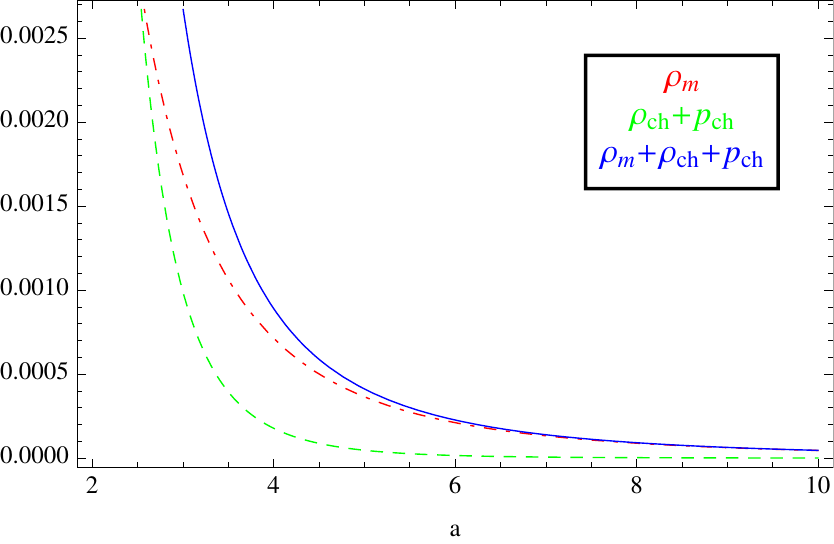}
 \end{minipage}
 \hspace{4 cm}
\begin{minipage}{0.3\textwidth}
\centering
 \includegraphics[width=8cm]{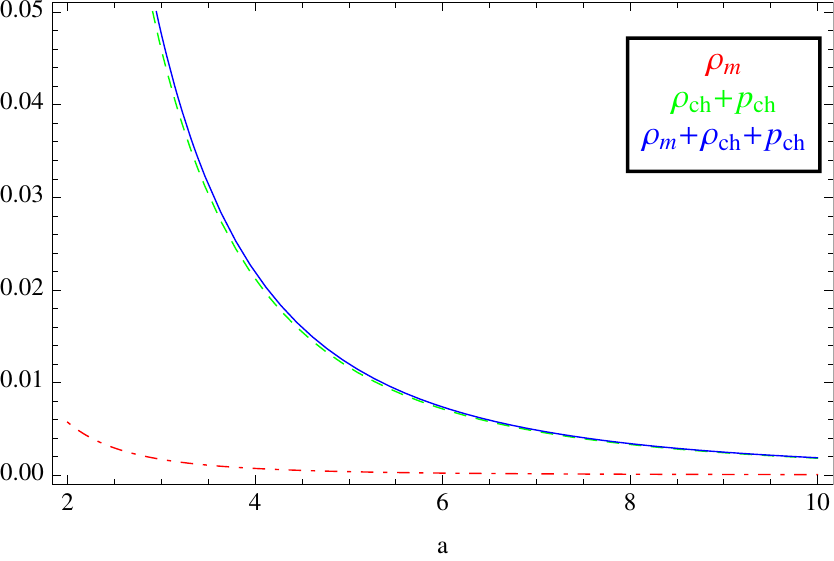}
  \end{minipage}
 \caption{Qualitative evolution of the energy densities and pressure $\rho_m$ (dot-dashed line), $\rho_{ch}+p_{ch} $ (dashed) and the sum of the two (solid) for different values of the parameters of the GBIG Chaplygin gas model. The left panel shows the evolution for the best-fit values by fitting the model with data from SNIA (gold sample). In the right panel all the observational data (SNIa, CMB and SDSS) are used in the fit. } \label{densityparameters}
\end{figure*}

\begin{table}
    \begin{tabular}{lccccccc}
       \hline 
  \text{Observations }& $\Omega_{r}$&$\Omega_{\alpha}$&$A_s$&$\Omega_{\Lambda}$&$\Omega_{ch}$&$\tilde{\alpha}$&$\chi^2_{\min}/N_{d.o.f.}$ \\ \hline
    \text{SNIa (gold sample)}&0.43&0.11&0.50&1.45&0.99&0.99&0.923  \\
    \text{SNIa(gold sample)+CMB+SDSS}&0.51&0.41&0.099&1.79&0.99&-0.98&0.992\\\hline
    \end{tabular}
\caption{Best fit values for the parameters of the GBIG Chaplygin model by fitting with SNIa gold sample and SNIa+CMB+SDSS. Cosmological parameters are defined as $\Omega_{r}=1/4rH_0^2$,  $\Omega_{\alpha}=8\alpha H_0^2/3$, $\Omega_{\Lambda}=\kappa_4^2\Lambda/3 H_0^2$, and $\Omega_{ch}=\rho_{ch0} \kappa_4^2/3 H_0^2$. $\Omega_m=\kappa_4^2\Lambda/3 H_0^2$ has been fixed to $\Omega_m=0.0456$, according to WMAP and BBN results \cite{nozari11}.}\label{fitpar}
\end{table}
Let us evaluate the sign of these derivatives: in order to fulfill the GSL the sum of the energy density and pressures of CDM and  Chaplygin must be positive. To estimate it we use the best fit results obtained in \cite{nozari11} whose model was tested using data from SNIa (gold sample), CMB (shift parameter) and SDSS (BAO peak). Qualitative different evolution and best-fit results are obtained for SNIa data and the joint  analysis of the total data set, as can be seen in Fig.\ref{densityparameters} and table \ref{fitpar}: the two panels plot the evolution of the energy densities and pressure multiplied by $3H_0^2/(8\pi G)$: $\rho_m$, $\rho_{ch}(1+w_{ch})$, and the sum of the two. On the left panel best fit values for the solely SNIa fit has been used and a slower redshift of the CDM can be appreciated. On the contrary, when SNIa+CMB+SDSS fit is taken into account, as displayed in the right panel, the Chaplygin gas redshifts more slowly. Then, in what follows we will test our thermodynamic hypothesis using both fits.

Moreover, Fig.\ref{densityparameters} shows that the term in  brackets is always positive which ensures the validity of the GSL.\\
The second derivative is negative, the minus sign given by the terms in squared brackets that, for SNIa fit tend to $-4$, while in the other case tend to $-4-3\tilde{\alpha}\simeq-1.06$: the term with the derivative of the horizon radius tends to zero, since the model tends to De Sitter, and the term involving energy densities tends to zero for the SNIa fitting and to $-3\tilde{\alpha}$  for the SNIa+CMB+SDSS one.

\subsection{Entropy of matter fields}
In this section we evaluate the role of the CDM entropy. We note that eqs.(\ref{Sdotdust})-(\ref{Sddotdust}) still apply.\\
Computing, as before, the ratio between CDM and horizon entropy derivatives one can see that for the best-fit values obtained just using the SNIa data
\begin{equation}
\frac{S'_m}{S'_A}\sim\frac{S''_m}{S''_A}\sim\text{const}
\end{equation}
so that eqs.(\ref{GSL})-(\ref{equilibrium}) as well as the constraint below still hold.\\
On the other hand, when computing the ratio in the second fit case, due to the fact that Chaplygin gas component redshifts slower than CDM, both first and second derivatives ratios tend to zero. In this case the horizon contribution prevails in the long run  ensuring that both the GLS is fulfilled and the equilibrium will be eventually approached. \\
As before, we can argue that only in the case that a dark energy component eventually dominates, the second law of  thermodynamics is fulfilled.

Finally, we evaluate the Chaplygin gas contribution to the entropy.
It yields
\begin{eqnarray}
S'_{ch}&=&4\pi\frac{ \tilde{r}_A^3(\rho_{ch}+p_{ch})}{T_{ch}a}\left[a\frac{\tilde{r}'_A}{\tilde{r}_A}-1\right],\label{Sdotch}\\
S''_{fi}&=&4\pi\frac{ \tilde{r}_A^2(\rho_{ch}+p_{ch})}{T_{ch} a}\left\{a \tilde{r}''_A+2a\frac{\tilde{r}'^2_A}{\tilde{r}_A}-6\tilde{r}'_A+4\frac{\tilde{r}_A}{a}-3\tilde{\alpha}A\rho^{-(\tilde{\alpha}+1)}\frac{\tilde{r}_A}{a}\left[a\frac{\tilde{r}'_A}{\tilde{r}_A}-1\right]-\frac{T'_{fi}}{T_{fi}}\tilde{r}_A\left[a\frac{\tilde{r}'_A}{\tilde{r}_A}-1\right]\right\}\label{Sddotch}.
\end{eqnarray}
The temperature evolution is found by solving $T'_{ch}=-3w_{ch}T_{ch}/a$. It turns to be
\begin{equation}
T_{ch}=T_{ch0}\left(1-A_s+A_s a^{3(1+\tilde{\alpha})}\right)^{1/(1+\tilde{\alpha})}.
\end{equation}
For both best-fit we find that 
\begin{equation}
\frac{S'_{ch}}{S'_A}\rightarrow 0\quad\text{and}\quad \frac{S''_{ch}}{S''_A}\rightarrow 0
\end{equation}
when $a\rightarrow \infty$. Thus, this contribution does not change previous results: a dark energy contribution must be present in the model and must dominate in the long run if the Universe is to eventually tend to thermodynamical equilibrium.
\section{Conclusions}\label{conclusions}

We have shown that a Gauss-Bonnet model with induced gravity corrections tends to thermodynamical equilibrium in the long run provided that at least one of the cosmological fluids corresponds to dark energy. The apparent horizon entropy is well-behaved, i.e. it is a growing and convex function of the scale factor, while the entropy of the fluids, either cold matter or radiation, is growing but diverging. Nevertheless, if a dark energy fluid ($w<0$) prevails in the long run, the overall entropy tends to a finite maximum value.\\
An explicit example is given by a Chaplygin gas component that behaves as a dynamical dark energy, smoothly crossing the phantom divide line and approaching the cosmological constant behaviour. The model under analysis is dominated by two energy components, CDM and Chaplygin gas. Best fit values for the parameters of the model were obtained in \cite{nozari11} and very different results are found when testing the model against SNIa data or SNIa+CMB+SDSS set. This implicitly translates into a different weight of the Chaplygin gas component in the long run thus altering the results on the entropy functions. When we use the best-fit values of the SNIa data alone, cold matter entropy evolves the same way as the horizon entropy, then in order to fulfill thermodynamical requirements the horizon radius must be larger than the Compton length of the cold matter particles. On the other hand, when we use SNIa+CMB+SDSS best fit results, since the Chaplygin gas redshifts more slowly, the overall entropy in the long run reduces to the horizon entropy that evolves toward equilibrium.

\acknowledgements
NR was funded by the Spanish Ministry of Science and Innovation through the "Subprograma Estancia J\'ovenes Doctores Extranjeros, Modalidad B", Ref: SB2009-0056. This research was partly supported by the Spanish Ministry of Science and Innovation under Grand FIS2009-13370-C02-01, and the "Direcci\'o de la Recerca de la Generalitat" under Grant 2009GR-00164.


\end{document}